\documentclass[twocolumn,aps,prl,superscriptaddress]{revtex4}

\usepackage{amsmath}
\usepackage{graphicx}
\usepackage{MnSymbol}

\begin{document}

\title{Two comments on adhesion}

\author{ A. Tiwari}
\affiliation{PGI-1, FZ J\"ulich, Germany, EU}
\affiliation{www.MultiscaleConsulting.com}
\author{J. Wang}
\affiliation{PGI-1, FZ J\"ulich, Germany, EU}
\affiliation{College of Science, Zhongyuan University of Technology, Zhengzhou 450007, China}
\author{ B.N.J. Persson}
\affiliation{PGI-1, FZ J\"ulich, Germany, EU}
\affiliation{www.MultiscaleConsulting.com}

\begin{abstract}
The adhesion paradox refers to the observation that for most solid objects no adhesion can be detected when they are separated from a state of molecular contact. The adhesion paradox results from surface roughness, and we present experimental and theoretical results which shows that adhesion in most cases is "killed" by the longest wavelength roughness.

Adhesion experiments between a human finger and a clean glass plate were carried out, and for a dry finger, no macroscopic adhesion occurred. We suggest that the observed decrease in the contact area with increasing shear force results from non-adhesive finger-glass contact mechanics, involving large deformations of a complex layered material.

\end{abstract}

\maketitle


{\bf 1 Introduction}

All solids have surface roughness extending over many decades in length scale, which has a large
influence on most tribology topics, such as adhesion, boundary and mixed lubrication and the leakage of seals.
Surface roughness is the main reason for why macroscopic solid objects usually do not adhere to each other, in spite of the strong force
fields which act at the atomic length scale between all atoms; this fact is referred to as the
adhesion paradox\cite{Kendall}. Strong adhesion between two macroscopic objects is observed only if both solids have very smooth surfaces,
and if at least one of the solids is elastically very soft. In Nature insects, lizards and tree frogs,
have ``learned'' (via natural selection) how to construct soft adhesion pads from stiff materials using hierarchical
building principles\cite{BoBio,BoSur}.

In this paper we will address two topics which recently have been discussed controversially 
in the literature\cite{Past2,PNAS,Cia1}. We first present
experimental and theory results showing that adhesion is ``killed'' mainly by the longest wavelength roughness. This is the
reason for why small particles may adhere strongly (agglomerate), while macroscopic solids of the same material may show no adhesion 
during approach or separation. We also present experimental results for adhesion between a human finger and a glass plate,
which is relevant for haptic applications\cite{Haptic,Finger1}.

        \vskip 0.3cm
        {\bf 2 Experimental}
	        \vskip 0.1cm
	{\bf Sandblasting and surface topography}
	
        We have sandblasted two polymethymethacrylate (PMMA) sheets with glass beads (spherical particles with smooth surfaces) 
	of diameter $\approx 10  \ {\rm \mu m}$ for a time ranging from $1-4$ minutes
	using $5-8$ bar air pressure. The topography measurements were performed with
	Mitutoyo Portable Surface Roughness Measurement device, Surftest SJ-410 
	with a diamond tip with the radius of curvature $R = 1 \ {\rm \mu m}$, and
	with the tip--substrate repulsive force $F_{\rm N}=0.75 \ {\rm mN}$.
	The lateral tip speed was $v=50 \ {\rm \mu m/s}$.
	
	From the the measured surface topography (line scans), $z=h(x)$, we calculated the 
	one-dimensional (1D) surface roughness power spectra defined by 
	$$C_{\rm 1D} (q) = {1\over 2 \pi} \int dx \ \langle h(x) h(0) \rangle e^{i q x}$$ 
	where $\langle .. \rangle$ stands for ensemble averaging. For surfaces with isotropic 
	roughness the two dimensional (2D) power spectrum $C(q)$ can be obtained directly from $C_{\rm 1D} (q)$
	as described elsewhere\cite{Nyak,CarbLor}. For randomly rough surfaces, all the (ensemble averaged)
	information about the surface is contained in  the power spectrum $C(q)$.
	For this reason the only information about the surface roughness which enter
	in (analytic) contact mechanics theories (with or without adhesion) is the function $C(q)$.
	
        \vskip 0.1cm
	{\bf Replicating roughness of PMMA on PDMS}
	
        The sandblasted PMMA surfaces was cleaned with distilled water 
        and then dried. We produced elastomer replicas of the two rough surfaces, and of a smooth PMMA surface,
        using Sylgard 184 Polydimethylsiloxane (PDMS) obtained from Dow Corning. 
        This elastomer is obtained from two liquid components,
        a pre-polymer base, B,  and a crosslinking agent, C. The two components can be mixed 
        in varying ratios to obtain desired elastic properties. For our purpose we prepared PDMS by choosing 
        C:B ratio of 1:10. We poured the PDMS fluid on the smooth and sandblasted PMMA surfaces and 
        kept it on a heated plate maintained at $70\,^{\circ}\mathrm{C}$ for 24 hrs. After this curing process, 
        we slowly removed the ($1 \ {\rm cm}$ thick) PDMS sheets from the PMMA surfaces.
        It has been shown in the past that PDMS can replicate roughness down to the nanoscale\cite{Julia, sab}.
	
        \vskip 0.1cm
	{\bf Adhesion measurement}
	
\begin{figure}[tbp]
		\includegraphics[width=0.47\textwidth,angle=0]{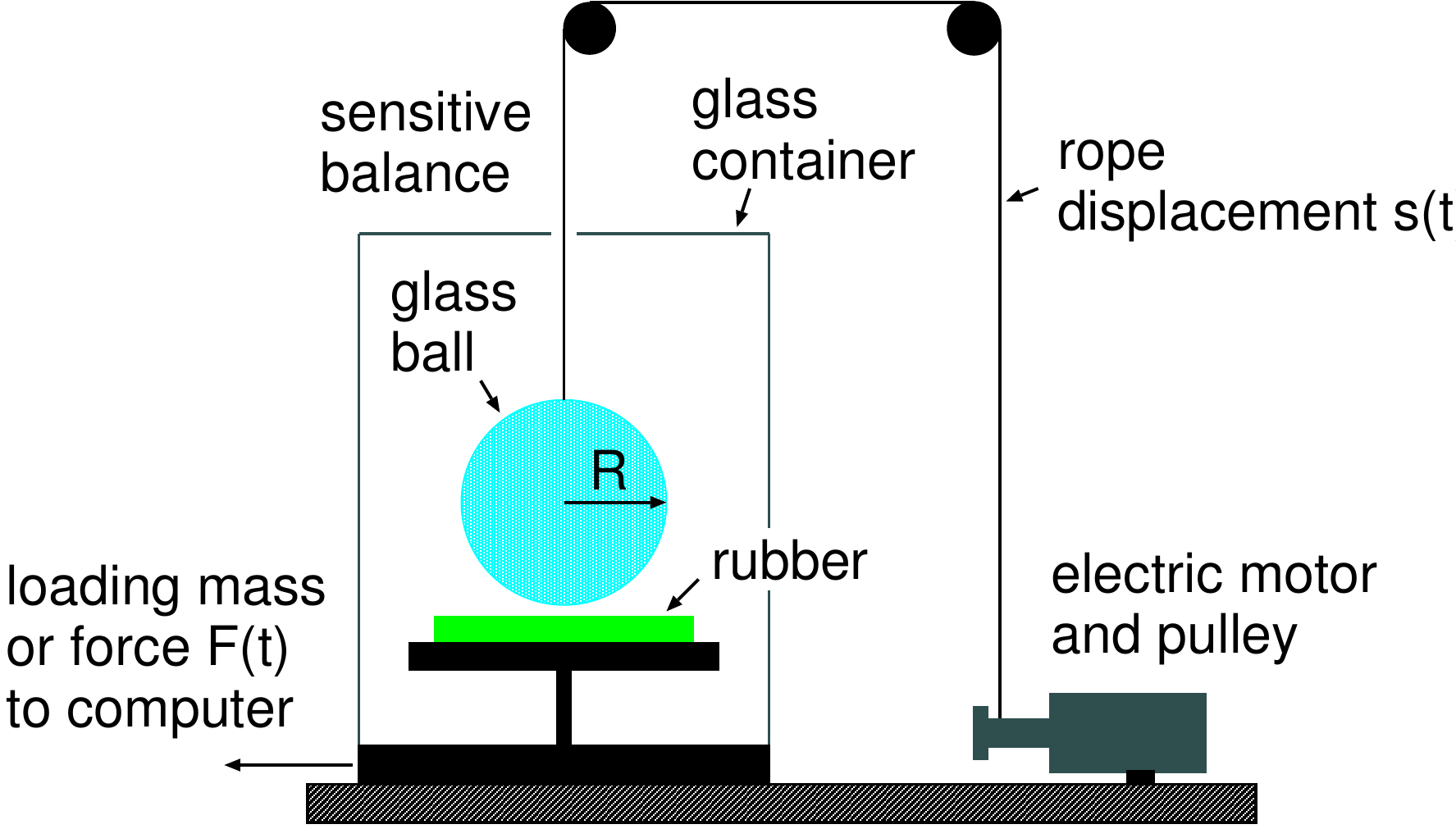}
		\caption{The J\"ulich experimental set-up for measuring adhesion.
		}
		\label{JuelichSetUp.ps}
	\end{figure}

        We have studied the adhesion interaction between a spherical glass ball (diameter $2R=2.5 \ {\rm cm}$) 
        and the PDMS rubber sheets. In the experiments we brought a glass ball into contact with a rubber 
        substrate as shown in Fig. \ref{JuelichSetUp.ps}. The rubber sample is positioned on a very accurate balance 
        (analytic balance produced by Mettler Toledo, model MS104TS/00),
	which has a sensitivity of $0.1 \ {\rm mg}$ (or $\approx 1 \ {\rm \mu N}$).
	After zeroing the scale of the instrument we can measure the force on the substrate
	as a function of time which is directly transferred to a computer at a rate of 10 data points per second.
	
	To move the glass ball up and down we have used an electric motor coiling up a nylon cord,
	which is attached to the glass ball. The pulling velocity as a function of time can be
	specified on a computer. In the experiments reported on below the glass ball is repeatedly
	moved up and down with the speed $25 \ {\rm \mu m/s}$, for up to $\sim 25$ contact cycles, 
        involving a measurement time of up to $20$ hours.
        
	The adhesion between human finger and smooth glass plate in dry state was studied in setup described above by application of
	$\approx$ 0.6 N force nominally on the smooth glass plate by finger and then slowly removing the finger away from the glass plate. The glass plate was cleaned with acetone and isoprapanol and the human finger of one of the authors was cleaned with soap water before each experiment. For adhesion in water a drop of water was placed on the glass plate and then the procedure same as the dried state was repeated.\\

\begin{figure}
\centering
\includegraphics[width=0.45\textwidth]{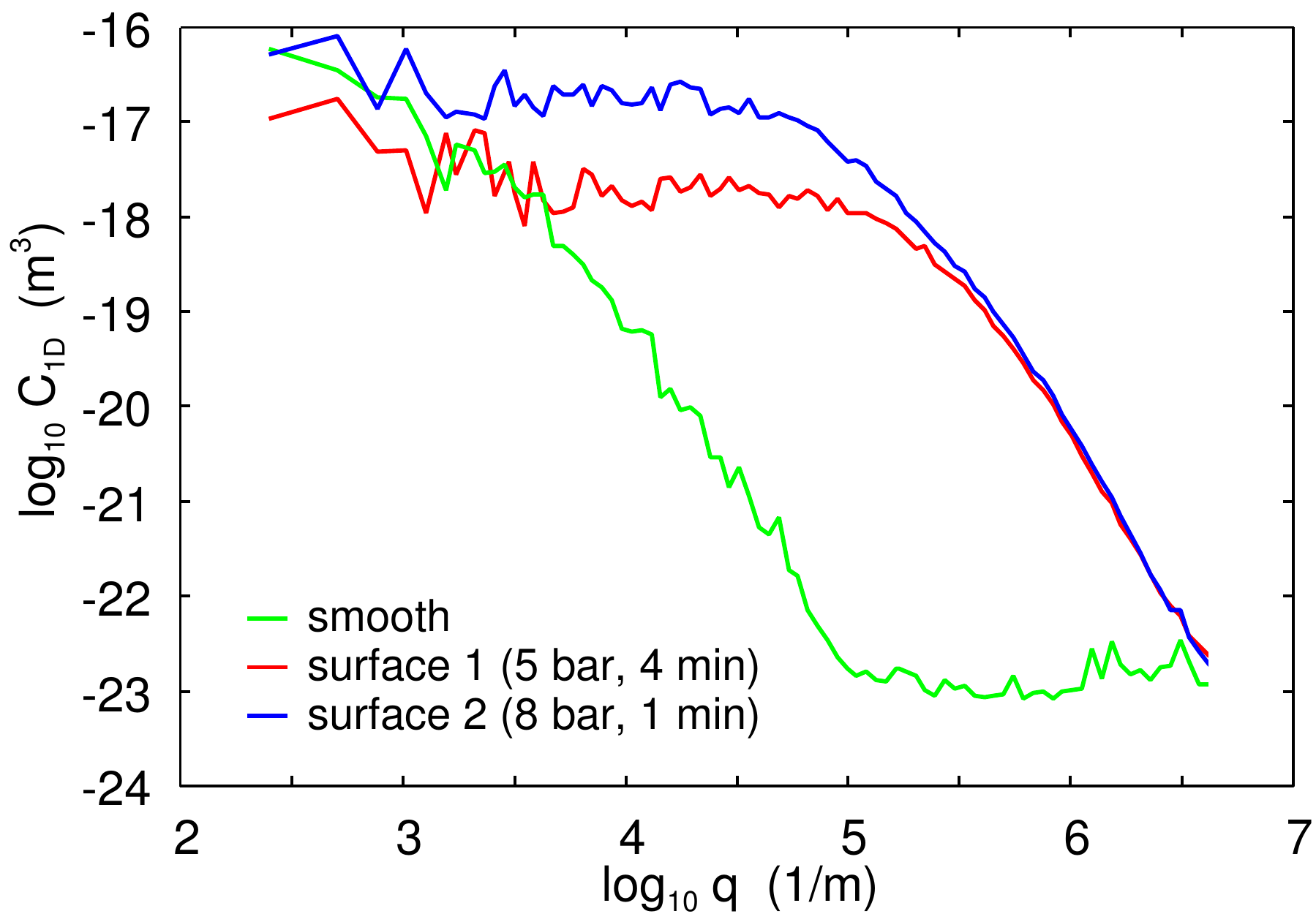}
\caption{\label{1logq.2logC1D.1.14.2.pdf}
The green, red and blue lines shows the wavenumber dependency of the
1D surface roughness power spectra for the smooth, sandblasted 1 and sandblasted 2
surfaces, respectively (log-log scale).
}
\end{figure}

\begin{figure}
\centering
\includegraphics[width=0.45\textwidth]{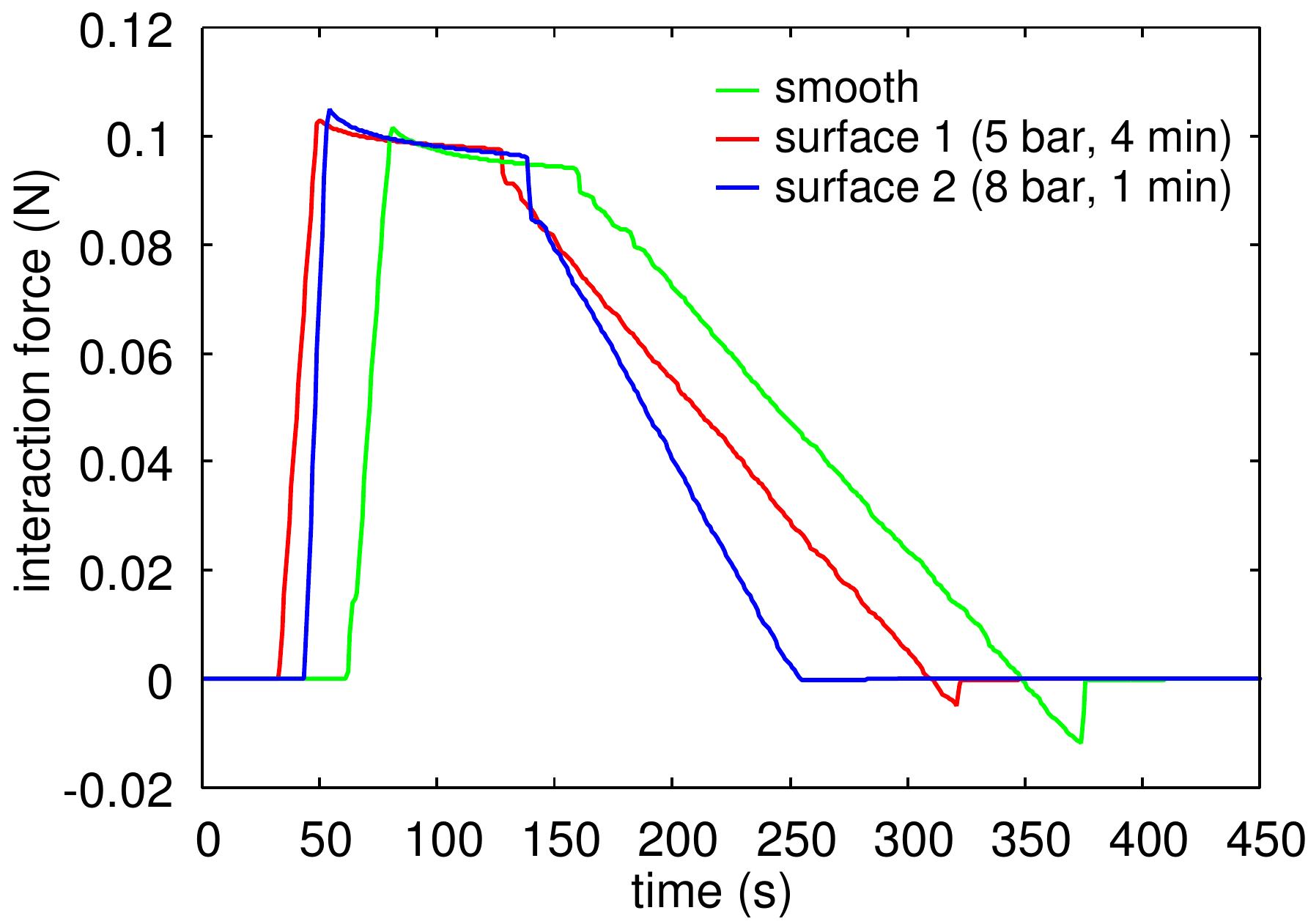}
\caption{\label{1time.2force.5bar.0bar.8bar.pdf}
The interaction force between the glass ball and the three PDMS surfaces.
The green, red and blue lines are for the smooth, sandblasted 1 and sandblasted 2 
surfaces, respectively. Note that no adhesion can be detected on approach, and for the
sandblasted 2 surface also not during pull-off (retraction).
}
\end{figure}

\begin{figure}
\centering
\includegraphics[width=0.45\textwidth]{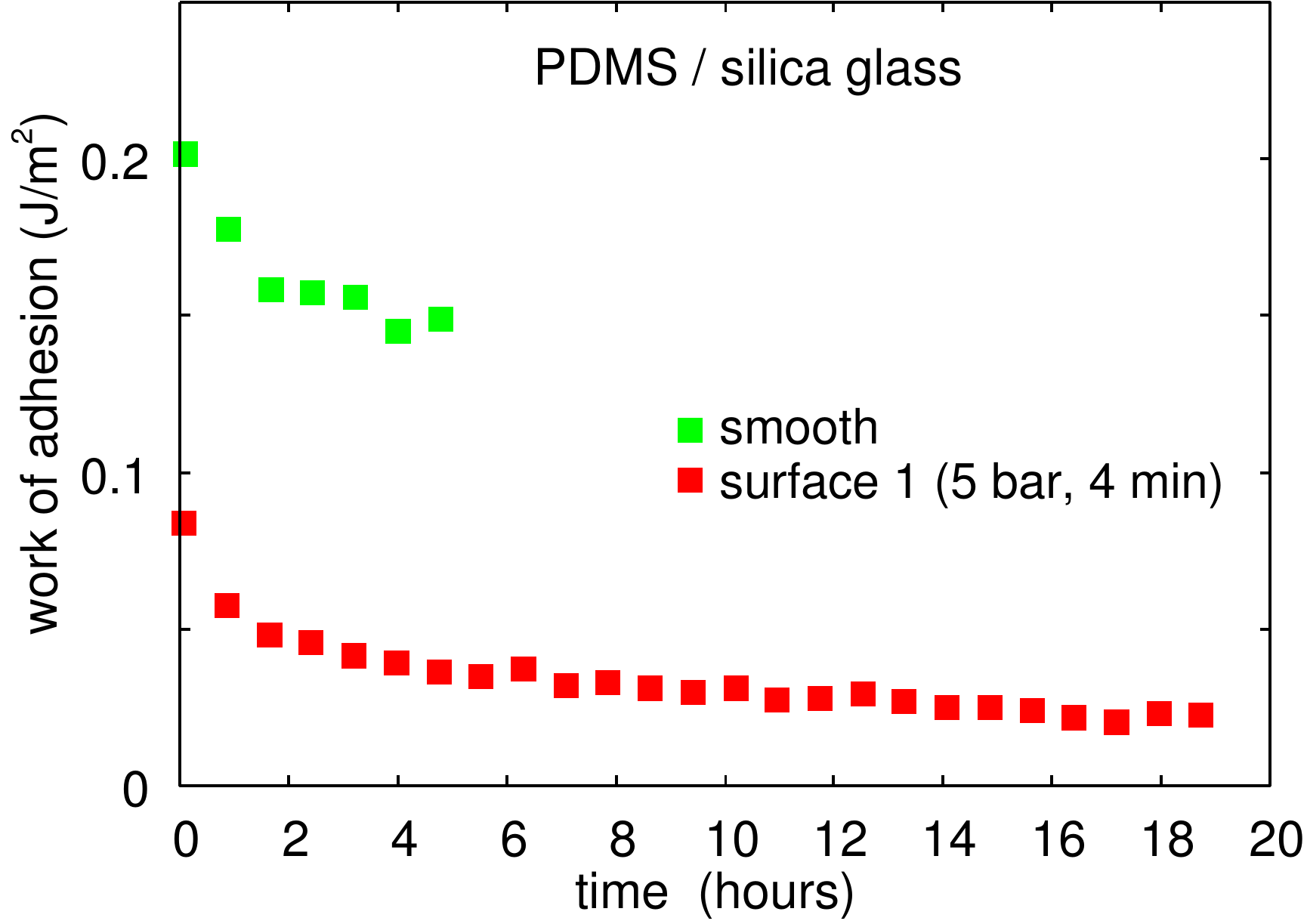}
\caption{\label{1time.2WorkAdhesion.Smooth.5bar.pdf}
The work of adhesion as a function of the number of contacts for the smooth PDMS surface (green squares)
and for the sandblasted surface 1 (red squares).
}
\end{figure}

\begin{figure}
\centering
\includegraphics[width=0.45\textwidth]{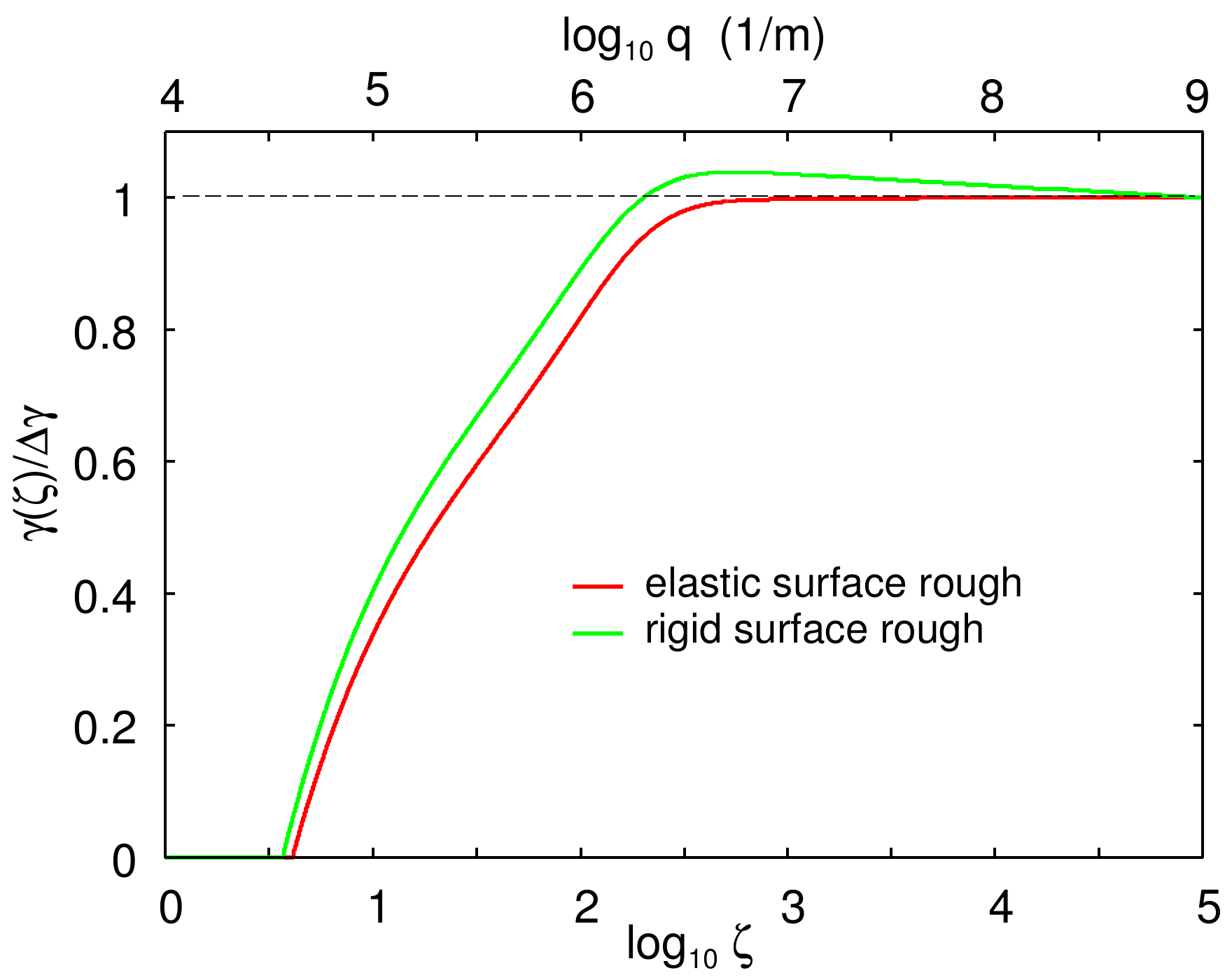}
\caption{\label{1logz.2w.red.rubberRough.green.hardRough.pdf}
The effective interfacial binding energy (per unit surface area), or work of adhesion,
as a function of the magnification (lower scale) or the wavenumber (upper scale). Note that $\gamma (\zeta)$
at the magnification $\zeta$ is the interfacial binding energy including only the roughness components with
$q > \zeta q_0$ (where $q_0$ is the smallest wavenumber). The red and green line is for the roughness on the
elastic solid and rigid solid, respectively.
In the calculation we used $\Delta \gamma =0.1 \ {\rm J/m^2}$ and $E^*=1.0 \ {\rm MPa}$.
}
\end{figure}

\begin{figure}
\centering
\includegraphics[width=0.45\textwidth]{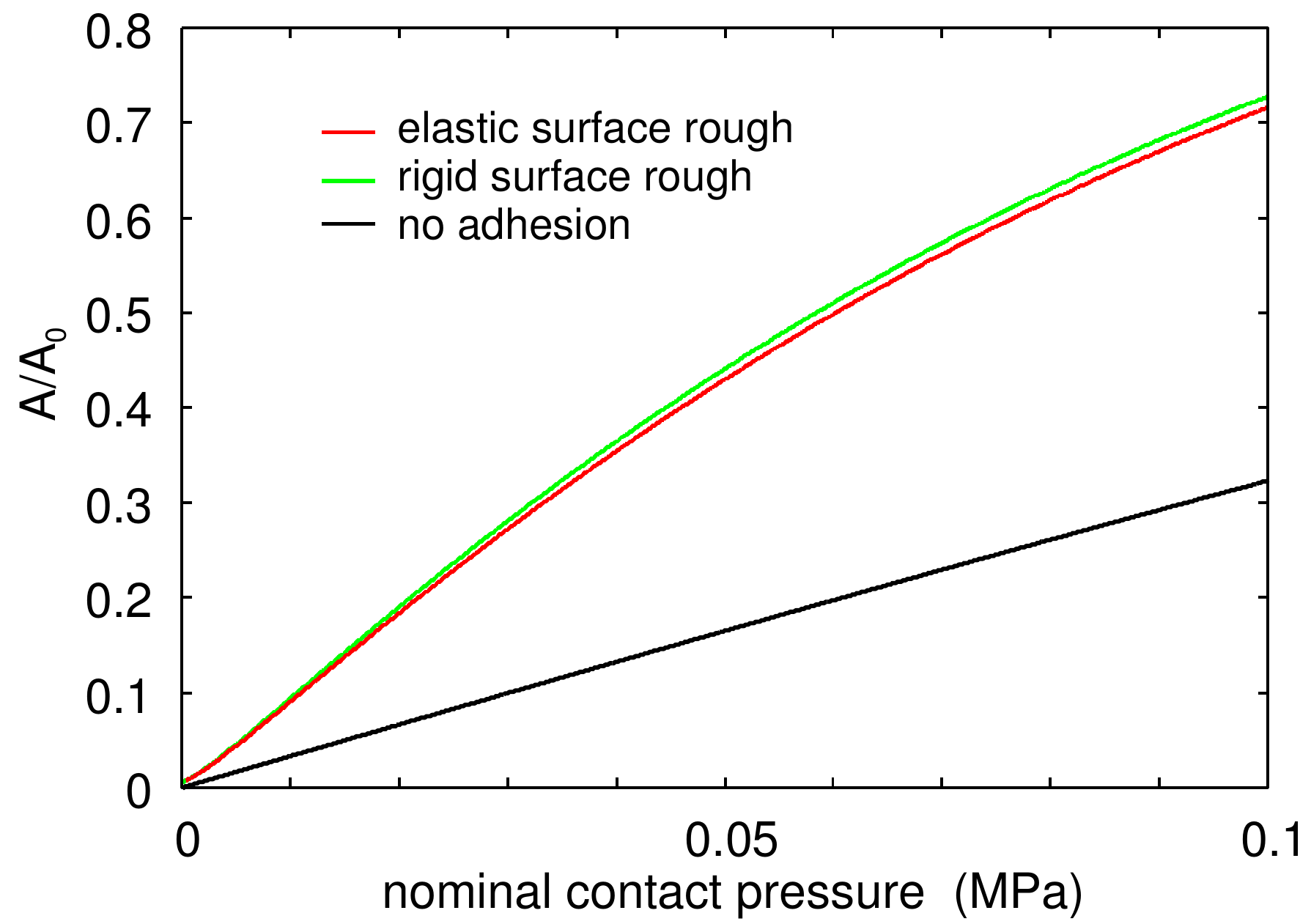}
\caption{\label{1pressure.2Area.red.rubberRough.green.hardRough.pdf}
The projected area of real contact $A$, normalized by the nominal contact area $A_0$,
as a function of the nominal applied (squeezing) pressure.
The red and green line is for the roughness on the
elastic solid and rigid solid, respectively. The black line is the result without adhesion.
}
\end{figure}

\begin{figure}
\centering
\includegraphics[width=0.45\textwidth]{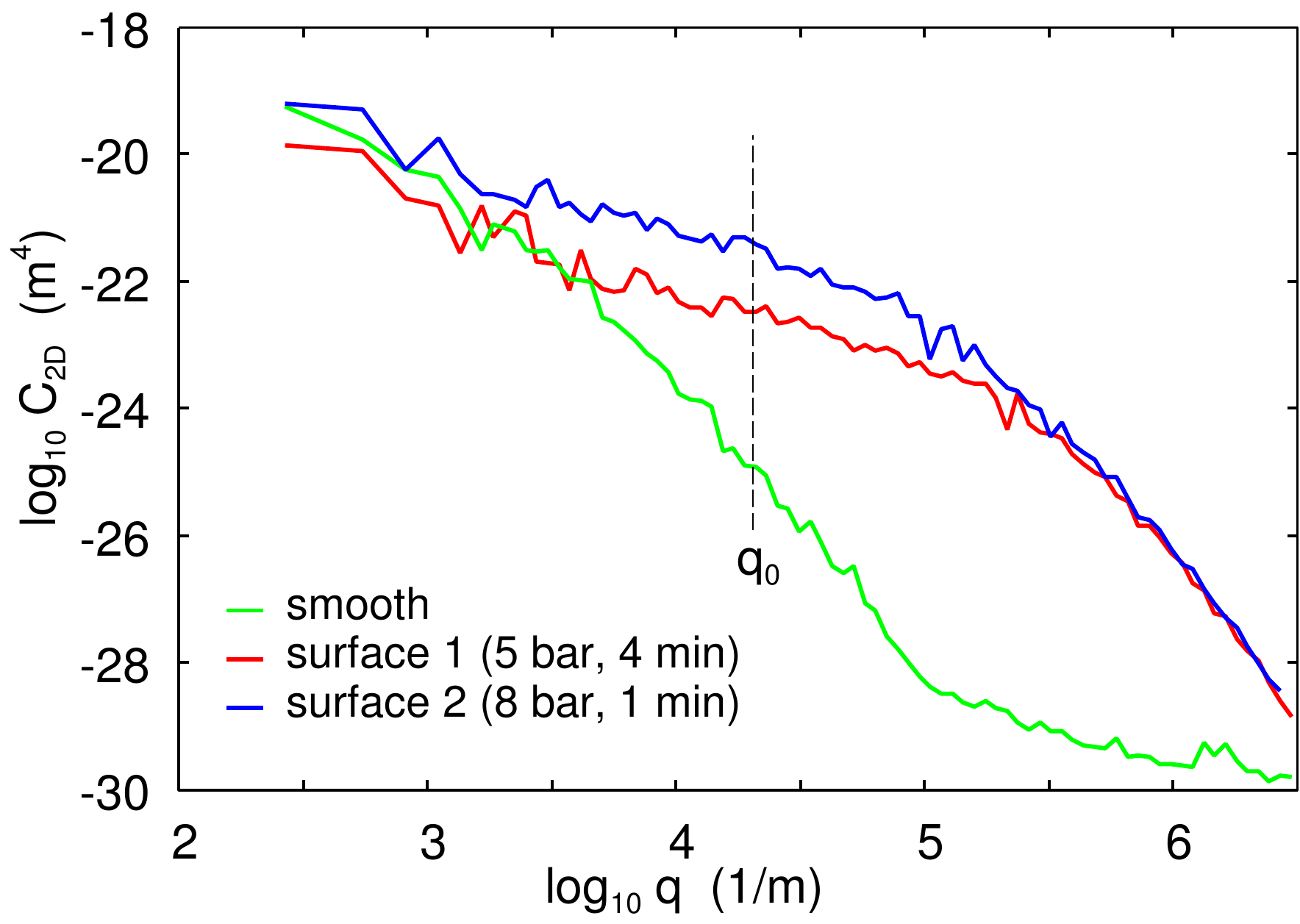}
\caption{\label{1logq.2logC2D.smooth.5bar.8bar.pdf}
The green, red and blue lines shows the wavenumber dependency of the
2D surface roughness power spectra for the smooth, sandblasted 1 and sandblasted 2
surfaces, respectively (log-log scale). The dashed vertical line indicate the lower
cut-off wavenumber used in the adhesion calculations.
}
\end{figure}

\begin{figure}
\centering
\includegraphics[width=0.45\textwidth]{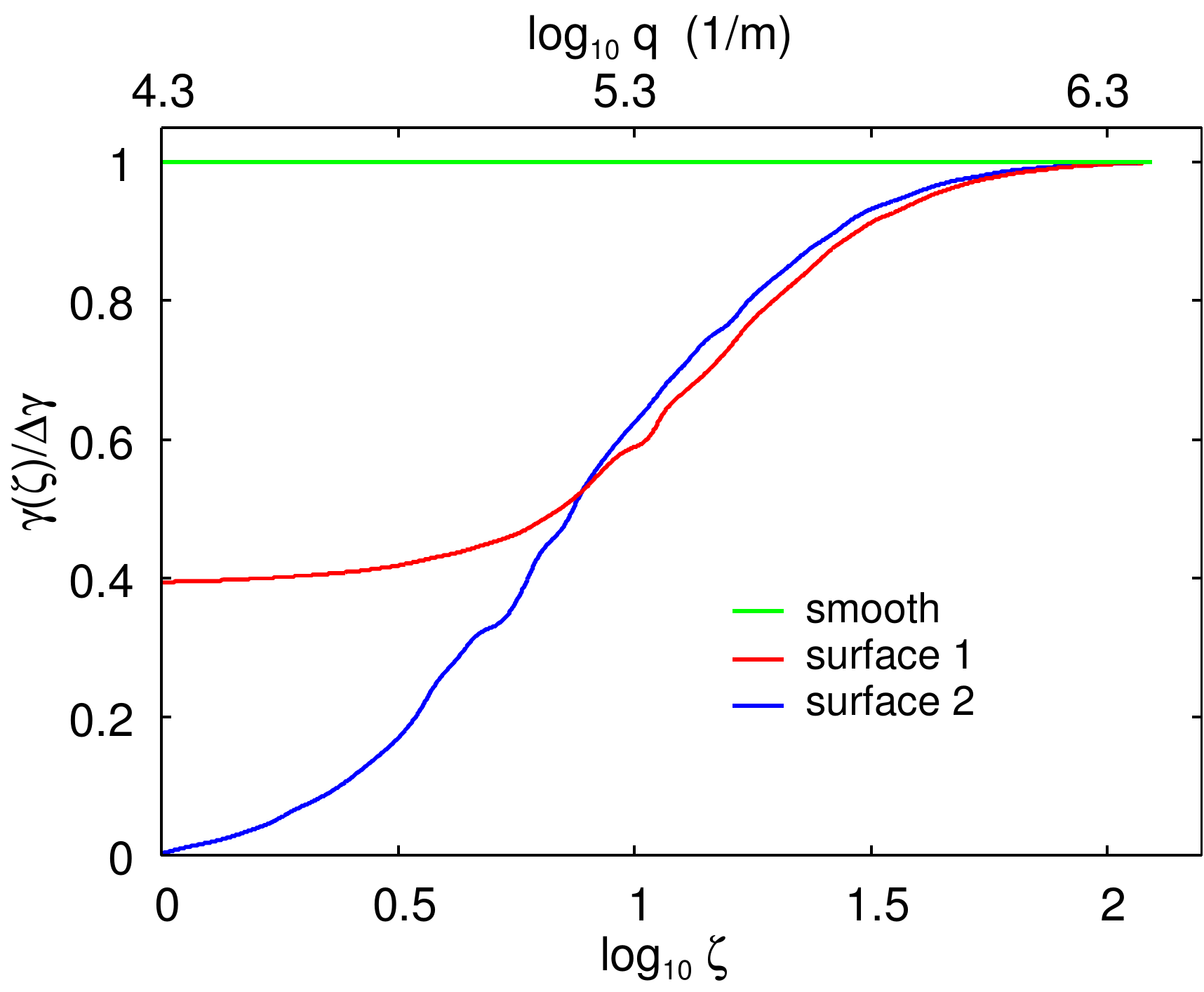}
\caption{\label{1logzeta.2ratio.work.pdf}
The effective interfacial binding energy (per unit surface area), or work of adhesion,
as a function of the magnification (lower scale) or the wavenumber (upper scale). Note that $\gamma (\zeta)$
at the magnification $\zeta$ is the interfacial binding energy including only the roughness components with
$q > \zeta q_0$ (where $q_0$ is the smallest wavenumber). The green, red and blue lines are for the 
smooth, sandblasted 1 and sandblasted 2 surfaces, respectively. 
In the calculation we used $\Delta \gamma =0.2 \ {\rm J/m^2}$ and $E=2.3 \ {\rm MPa}$, $\nu = 0.5$.
}
\end{figure}

\vskip 0.5cm
{\bf 3 Role of surface roughness at different length scales on adhesion}\\
Due to the surface roughness, which exist on all solid objects, adhesion between macroscopic solid bodies is 
usually not observed. The reason is that to make contact the solids must deform at the contacting interface,
and this stores up elastic energy at the interface which is given back during separation of the solids, and
help to break the atomic bonds at the contacting interface. The elastic energy required to bend the surface of a solid
(with the Young's elastic modulus $E$)
so that it fills out a cavity of width $\lambda$ and depth $h<<\lambda$ is of order $E_{\rm el} \approx 
\lambda^3 E (h/\lambda )^2 = \lambda h^2 E$, while the gain in adhesion energy is $E_{\rm ad} \approx \lambda^2 \Delta \gamma$,
where $\Delta \gamma$ is the work of adhesion (the energy per unit surface area to separate two 
flat surfaces of the two materials involved). Here we have used that the elastic strain $\epsilon \approx h/\lambda$
and that the deformation field extend over a volume $\approx \lambda^3$. The ratio $E_{\rm el}/E_{\rm ad} = (E/\Delta \gamma) (h^2/\lambda)$.
If we consider cavities of different size but with the same ratio $h/\lambda$ then $E_{\rm el}/E_{\rm ad} \sim \lambda$ i.e.
the elastic energy will dominate over the adhesive energy at large enough length scales $\lambda$. 
This is the basic reason adhesion is not observed
in most cases for macroscopic solids even if adhesion may be strong for small solid 
objects (e.g. nanoparticles) of the same material\cite{Per1,AviDoroBo1,Past1}.
It is the reason why small particles may agglomerate into bigger particles, while macroscopic bodies made from the same materials may
not adhere at all. Here we will present experimental results and calculations illustrating this fundamental conclusion.

\vskip 0.1cm
{\bf 3.1 Experimental results}

We have produced (nearly) randomly rough Polymethylmethacrylat (PMMA) surfaces using sandblasting.
From theories of growth (or here erosion) one expect the root-mean-square (rms) roughness to increase continuously with the
sandblasting time, with a roll-off in the surface power spectrum which is moving to longer
wavelength (shorter wavenumber) with increasing sandblasting time\cite{Fractal,BoTL}. Thus theory predicts that the short wavelength roughness
is independent of the sandblasting time, while more longer-wavelength roughness is added with increasing sandblasting time.
We have observed the same effect by varying the kinetic energy of the sandblasting particles. 

The green, red and blue lines in Fig. \ref{1logq.2logC1D.1.14.2.pdf}
shows the wavenumber dependency of the
1D surface roughness power spectra for a smooth PMMA surface, and of two sandblasted surfaces 
(denoted 1 and 2), respectively (log-log scale).
The surface 1 was sandblasted for 4 minutes using 5 bar air pressure, and the surface 2 for 1 minute
using 8 bar air pressure. Note that the large wavenumber (short wavelength) power spectra of the two sandblasted
surfaces are the same while for small wavenumber the surface 2 has a larger power spectrum, which is reflected in the
rms-roughness amplitude which is $0.78 \ {\rm \mu m}$ and $1.73 \ {\rm \mu m}$ 
for surface 1 and 2, respectively. The rms-slope is determined mainly by
the short wavelength roughness and are nearly the same (0.18 and 0.22, respectively).  
The rms slope of the smooth surface is much smaller, 0.04.  

Using the adhesion set-up described in Sec. 2 we have measured the pull-off force between
the glass ball and the smooth and rough (surface 1 and 2) PDMS rubber surfaces.  
Fig. \ref{1time.2force.5bar.0bar.8bar.pdf}
shows the interaction force between the glass ball and the three PDMS surfaces during one contact.
The green, red and blue lines are for the smooth, sandblasted 1 and sandblasted 2 
surfaces, respectively. Note that no adhesion can be detected on approach for any of the surfaces, 
and for the sandblasted 2 surface also not during pull-off (retraction). That is, the additional
long-wavelength roughness of the surface 2 as compared to the surface 1, has killed the
macroscopic adhesion, while surface 1 shows a pull-off force roughly half as large as for the
smooth PDMS surface.

Using the Johnson-Kendall-Roberts (JKR) theory, from the pull-off force $F_{\rm c}$ we can obtain the work of adhesion
$$F_{\rm c} = {3\pi \over 2} \gamma R$$
In Fig. \ref{1time.2WorkAdhesion.Smooth.5bar.pdf}
we show the work of adhesion $\gamma$ to separate the surfaces 
as a function of the number of contacts for the smooth PDMS surface (green squares)
and for the sandblasted surface 1 (red squares). For surface 2 the work of adhesion to separate the
surfaces vanish. The work of adhesion during approach (contact formation) vanish for all three surfaces,
i.e., strong contact hysteresis occur in all cases. Note also that the
work of adhesion to separate the surfaces drops with the number of contacts, which we attribute to transfer of uncrosslinked
molecules from the PDMS to the glass ball, which has been observed also for other rubber compounds\cite{AviDoroBo,AviDoroBo1,AviDoroBo2}.

\vskip 0.1cm
{\bf 3.2 Theory results}

The two rough PDMS surfaces used above have large roll-off regions, which have a small influence on
the adhesion. We therefore first illustrate the role of different length scales on adhesion with a
case without a roll-off region. We consider a self-affine fractal surface with the fractal dimension 2
which imply that the ratio between the amplitude and the wavelength 
is the same independent of the wavelength of the roughness component. Thus on a log-log scale the
2D power spectra as a function of the wavenumber is a straight line with the slope $-4$ (see Ref. \cite{BoSur}). 
We assume the small and the large
cut-off wavenumbers $q_0=10^4 \ {\rm m}^{-1}$ and $q_1=10^9 \ {\rm m}^{-1}$. The surface has the
rms-roughness $10 \ {\rm \mu m}$ and the rms-slope $0.48$.

Fig. \ref{1logz.2w.red.rubberRough.green.hardRough.pdf}
shows the effective interfacial binding energy (per unit surface area), or work of adhesion,
as a function of the magnification (lower scale) or the wavenumber (upper scale). Note that $\gamma (\zeta)$
at the magnification $\zeta$ is the interfacial binding energy including only the roughness components with
$q > \zeta q_0$ (where $q_0$ is the smallest wavenumber). The red and green line is for the roughness on the
elastic solid, and on the rigid solid, respectively. In the present case $\gamma (\zeta)$ vanish before reaching the
$\zeta = 1$, i.e., there is no macroscopic pull-off force for either case. Note that the drop in $\gamma (\zeta)$
is due to the longer wavelength part of the roughness spectra. In fact, for the case where the roughness occur on the
rigid surface, the short wavelength part of the roughness spectra enhances the work of adhesion $\gamma (\zeta)$
for large $\zeta$. This effect is due to the increase in the surface area (we have assumed that the interfacial 
binding energy per unit surface area is unchanged by the increase in the surface area, which may hold for rubber-like
materials as they have a thin surface layer with liquid-like mobility).

Since there is no macroscopic adhesion, $\gamma (1)=0$, the contact area will vanish continuously as the applied
nominal contact pressure approach zero. This is shown in Fig.
\ref{1pressure.2Area.red.rubberRough.green.hardRough.pdf}. The figure shows
the projected area of real contact $A$, normalized by the nominal contact area $A_0$,
as a function of the nominal applied (squeezing) pressure.
The red and green line is for the roughness on the
elastic solid and on the rigid solid, respectively.
When macroscopic adhesion occur, i.e., $\gamma (1)>0$, the area of real contact is non-zero also when
the applied pressure vanish. See Ref. \cite{Amont,BoEPJE,BoMic} for results illustrating this.

Note that even if the macroscopic adhesion vanish (no pull-off force), the area of real contact is
increased by the adhesion. This implies, for example, that the adhesive interaction
will increase the sliding friction force even if no adhesion can be detected in a pull-off experiment. 

Next, let us consider the work of adhesion between the silica glass ball and the smooth and rough PDMS surfaces studied
above. From the 1D surface roughness power spectra shown in Fig. \ref{1logq.2logC1D.1.14.2.pdf} 
we first calculated the 2D surface roughness power
spectra shown in Fig. \ref{1logq.2logC2D.smooth.5bar.8bar.pdf}. Using the Persson contact mechanics theory\cite{BoEPJE}, 
in Fig. \ref{1logzeta.2ratio.work.pdf} we show the calculated effective interfacial binding energy (per unit surface area), or work of adhesion,
as a function of the magnification (lower scale) or the wavenumber (upper scale). 
In the calculation we have used the low wavenumber cut-off $q_0$ indicated by the dashed vertical line in the figure.
We have choosen $q_0 \approx 2 \pi /r_0$, where $r_0\approx 0.38 \ {\rm mm}$ is the JKR radius of the circular 
contact region at the point of snap-off
for the PDMS surface 1. The green, red and blue lines are for the 
smooth, sandblasted 1 and sandblasted 2 surfaces, respectively. In agreement with the experiments, the surface 2
has vanishing macroscopic ($\zeta = 1$) interfacial binding energy, while the surface 2 has $\gamma(1)/\Delta \gamma \approx
0.4$ close to the observed value for the first contact $\approx 0.083/0.201 \approx 0.413$
(see Fig. \ref{1time.2WorkAdhesion.Smooth.5bar.pdf}). For the smooth surface the theory
predict that the measured work of adhesion $\approx 0.2 \ {\rm J/m^2}$ is not influenced (reduced) by the roughness on the
PDMS surface.

\begin{figure}
\centering
\includegraphics[width=0.45\textwidth]{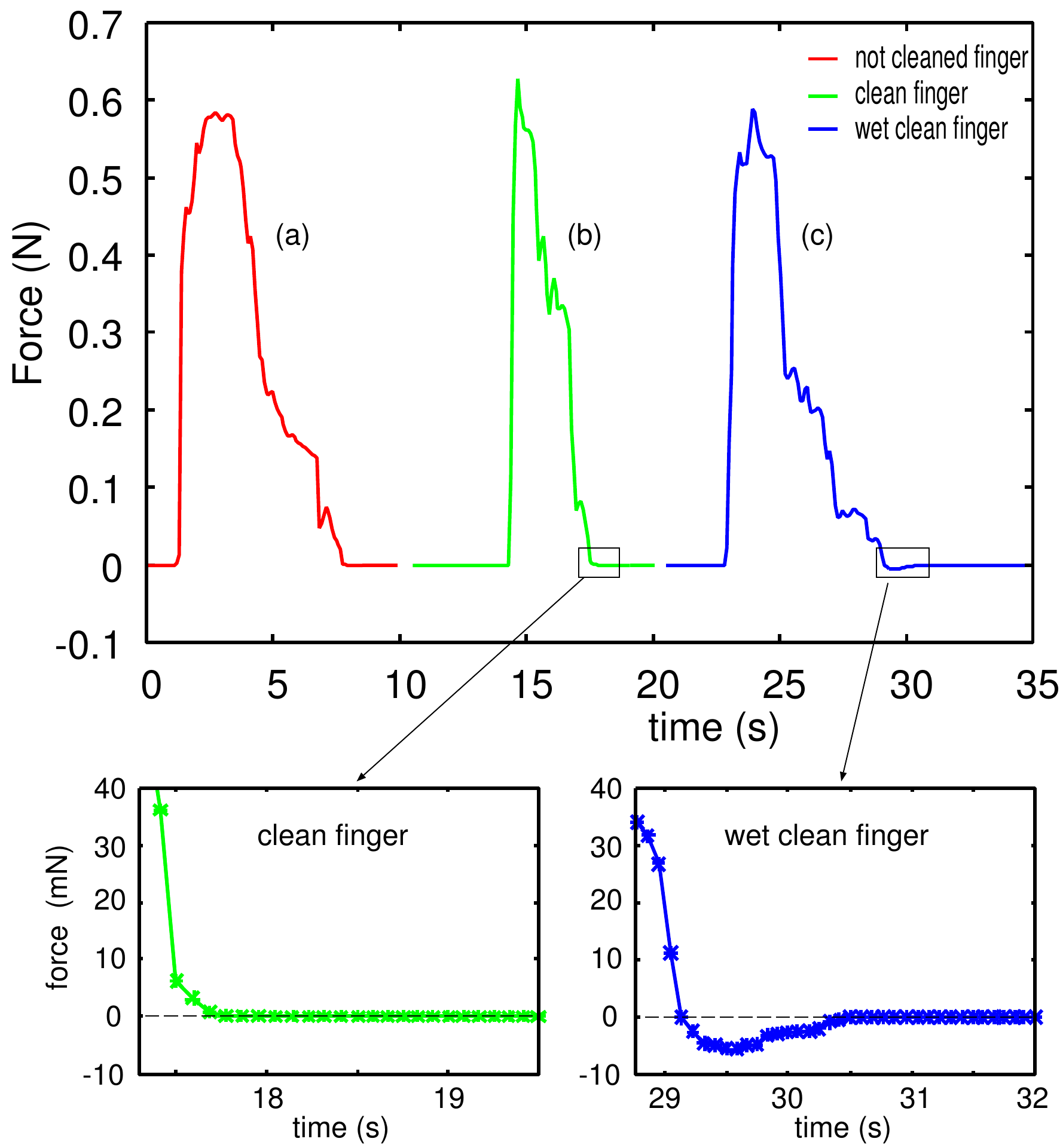}
\caption{\label{fingeradhesion}
The interaction force between a human finger and a dry glass plate cleaned by acetone and isopropanol.
Case (a) (red curve) is for a not cleaned finger,
(b) (green curve) for a finger cleaned with soap water and (c) (blue curve) for a clean wet finger.
In case (a) and (b) no (macroscopic) adhesion is observed, while in case (c) do we observe
adhesion with a pull-off force $F_{\rm c} \approx 5.5 \ {\rm mN}$.}
\end{figure}

\vskip 0.3cm
{\bf 4 Finger-glass adhesion experiments}

Several recent experimental studies have shown that when a tangential force is applied to a  human finger squeezed against a
flat glass surface, the glass-finger nominal contact area decreases\cite{PNAS,Finger1}. This has been tentatively explained using the
adhesion theory described in Ref. \cite{Cia}. However, we have performed adhesion experiments
for a finger in contact with a glass plate, and for a dry finger we do not observe any macroscopic adhesion so the explanation
proposed in Ref. \cite{Cia} cannot explain the observed
decrease in the contact area with increasing tangential force.

Fig. \ref{fingeradhesion}
shows the interaction force between a human finger and a dry glass plate cleaned by acetone and isopropanol.
Case (a) (red curve) is for a not cleaned finger,
(b) (green curve) for a finger cleaned with soap water and (c) (blue curve) for a clean wet finger.
In case (a) and (b) no (macroscopic) adhesion is observed, while in case (c) do we observe
adhesion with a pull-off force $F_{\rm c} \approx 5.5 \ {\rm mN}$.
This is similar to what is expected if a capillary bridge is formed between the glass surface and the finger.
Thus for a thick water film $F_{\rm c} \approx  4 \pi R \gamma_{\rm w}$,
where the water surface tension $ \gamma_{\rm w} \approx 0.07 \ {\rm J/m^2}$ and $R$ is the radius of curvature of the finger.
If we use $R\approx 0.7 \ {\rm cm}$ we obtain the observed pull-off force.
However, the pull-off force depends on the volume of water on the finger and if the water volume is too small (less then $\sim 1 \ {\rm mm}^3$)
no adhesion is observed which we interpret as resulting from the skin surface roughness and the elastic rebound of the deformed skin.

We believe that the reduction in the contact area observed for the human finger with increasing lateral force
is due to the complex inhomogeneous (layered) nature of the finger and to the large deformations involved.
It is also possible that the superposition of the normal and parallel deformation fields assumed in most analytic treatments
is not accurate enough when the parallel deformations becomes large and coupling effects becomes important.
This conclusion is supported by finite element calculations performed by Mergel et al\cite{Merg1} and more recently 
by Lengiewicz et al \cite{a1}(see also \cite{Merg2} and \cite{Julia}), which shows that even without adhesion there
is a reduction in the contact area between an elastic cylinder and a flat surface as a tangential force is applied to the cylinder.

\vskip 0.3cm
{\bf 5 Summary and conclusion}

In this article we studied two aspects of adhesion which shed light 
into recent debates in contact mechanics (with adhesion)\cite{Cia1,stick,Cia}. 
First, we discussed the adhesion paradox, the fact that adhesion is usually not observed
at macroscopic length scales. We presented experimental results and theoretical
calculations which showed that adhesion in most cases is ``killed'' by
the long-wavelength part of the roughness spectrum. 
Secondly, results of adhesion experiments between a human finger and flat smooth glass 
surface was presented. We found that there was no macroscopic adhesion between these contacting 
pairs in the dry state. Based on this result, we suggest that the decrease in the contact 
area as reported in the literature\cite{PNAS,Finger1, PRL} results from 
non-adhesive contact mechanics, involving large deformations of complex layered material.

\vskip 0.3cm
{\bf Acknowledgments}
J. Wang would like to thank scholarship from China Scholarship Council (CSC) and funding by 
National Natural Science Foundation of China (NSFC): grant number U1604131.

\end{document}